\documentclass[journal,12pt,onecolumn,draftclsnofoot]{IEEEtran}
\usepackage{amsmath}
\usepackage{amssymb}
\usepackage{graphicx}
\usepackage{color}
\usepackage{stackengine}
\usepackage{lineno}
\usepackage{cite}
\usepackage{lettrine}
\usepackage{bm,upgreek}
\renewcommand\IEEEkeywordsname{Index Terms}

\DeclareGraphicsExtensions{.pdf,.png,.jpg}

\IEEEoverridecommandlockouts

\begin{document}

\title{\LARGE \textit{M}-ary Orthogonal Chirp Modulation for Coherent and Non-coherent Underwater Acoustic Communications \vspace{-2pt}}
\author{Song-Wen Huang and Dimitris A. Pados

\thanks{S.-W. Huang is with Velodyne LiDAR, Inc., San Jose, CA 95138, USA. D. A. Pados is with the Department of Computer and Electrical Engineering and Computer Science, Florida Atlantic University, Boca Raton, FL 33431, USA. (e-mail: songwenh@buffalo.edu; dpados@fau.edu)}

\vspace{-16pt}
}

\maketitle

\begin{abstract}
We propose an orthogonal chirp waveform design for underwater acoustic (UW-A) communications and analyze the cross-correlation characteristics of orthogonal chirp waveforms in coherent and non-coherent detections. We consider information symbols are carried over proposed $\bm{M}$-ary orthogonal chirp waveforms for UW-A transmissions. Moreover, we develop a coherent and an optimal non-coherent receivers based on proposed chirp waveforms. Explicit derivations include closed-form expressions for cross-correlation coefficients, and theoretical bit-error-rate (BER) of coherent and non-coherent receivers. Performance of $\bm{M}$-ary orthogonal chirp waveforms is evaluated in water tank experiments. Therefore, we have demonstrated the effectiveness of proposed $\bm{M}$-ary orthogonal chirp modulation in UW-A multipath fading channel.
\end{abstract}


\IEEEpeerreviewmaketitle

\vspace{-8pt}
\section{Introduction}
\label{s:1}

Underwater acoustic (UW-A) communication is a field of rapid growing for numerous applications, including environmental monitoring, offshore remote control for resource exploration, underwater communications and networks, early warning for disaster and pollution control, and underwater surveillance \cite{melodia13,demirors15,akyildiz04,stojanovic96}. The UW-A channel is still suffered from harsh challenges, such as high attenuation, multipath propagation delays, Doppler spread, and range-dependent bandwidth, that influence technical advances notoriously \cite{stojanovic96}.

Chirp signals have been utilized extensively in radar and sonar applications \cite{fitzgerald74,wang15} primarily due to their robustness to multipath effect, Doppler spread, and superior correlation characteristics. Applications of UW-A communications include linear chirp signals for packet synchronization and channel estimation \cite{wu12}, reliable UW-A communications \cite{he09}, Doppler estimation and compensation \cite{sharif00}, and feedback link communications \cite{demirors14}.

Nonetheless, linear chirps are not rigorously orthogonal to each other \cite{springer00,huang17}. To solve this issue, some researchers adopt numerous kinds of chirp-based orthogonal transforms, such as fractional Fourier transform \cite{en14,solyman12}, and Fresnel transform \cite{ouyang16} for providing reliable multiuser and multicarrier communications. However, the drawbacks of these chirp-based transforms are of high computational overhead and demand complex hardware implementations. Therefore, there is still of demand and interest to develop orthogonal chirp waveforms, coherent and non-coherent receivers in UW-A communications.

In this paper, we consider a single-input single-output (SISO) system in UW-A communications that information symbols are carried over $M$-ary orthogonal chirp waveforms, termed as Orthogonal Chirp Keying (OCK). We develop a coherent receiver capable of processing multipath-affected signals and an optimal non-coherent detector \cite{Proakis07}. The orthogonality of chirp waveforms is analyzed, and close-form equations of theoretical bit-error-rate (BER) are derived explicitly. The channel state information (CSI) of the coherent receiver is estimated as complex coefficients using known pseudorandom noise (PN) training sequences. On the other hand, we can design an optimal non-coherent receiver based on proposed orthogonal chirp waveforms without any channel estimation. Then, proposed $M$-ary orthogonal chirp modulation is implemented with in-house built underwater acoustic modems and BER performance is evaluated in indoor tank experiments. 

The rest of the paper is organized as follows. Section \ref{s:2} addresses the system model. Receiver designs of coherent and non-coherent receivers are introduced in Section \ref{s:3}. Orthogonal chirp waveform design and analysis are presented in Section \ref{s:4}. Experimental results are demonstrated in Section \ref{s:5}. Finally, several concluding remarks are drawn in Section \ref{s:6}.

\vspace{-6pt}

\section{System Model}
\label{s:2}

We consider an underwater communication system that information symbols are carried over $M$-ary chirp waveforms in UW-A channel. Specifically, the $m$-th linear chirp waveform is represented as
\begin{align}
\psi_m(t) \triangleq \sqrt{\frac{1}{T}} e^{j(2\pi (f_0+ m\Delta f) t + \pi \mu t^2)},~0\leq t \leq T
\label{eq:1}
\end{align}
where $T$ is the symbol duration, $f_0$ is the initial frequency of the $0$-th ($m = 0$) chirp waveform, $\mu \triangleq \frac{B_c}{T}$ is the rate of frequency sweeping, $B_c$ is the bandwidth of the chirp waveform, $\Delta f$ is frequency spacing between chirp waveforms, and $m = 0,\cdots,M-1$. For the chirp rate, $\mu > 0$ is the up-chirp signal (increasing frequency with respect to time), while $\mu < 0$ is the down-chirp signal (decreasing frequency with respect to time). $f_0 + m\Delta f$ determines the initial frequency of the $m$-th chirp waveform. 

Therefore, the transmitted signals in passband are represented as
\begin{align}
x(t) = \sqrt{E}\sum_{n=0}^{N-1} \psi^{(n)}(t- nT)e^{j2\pi f_c t}
\label{eq:2}
\end{align}
where $E >0$ denotes transmitted energy per symbol, $N$ is the number of symbols, $\psi^{(n)}(t) \in \{\psi_m(t)\}_{m=0}^{M-1}$ is $n$-th transmitted symbol, and $f_c$ is the carrier frequency. 

In channel model, paths are assumed to be independent and time-variant. The multipath channel model is written as
\begin{align}
h(t) \triangleq \sum_{p=0}^{P-1} h_p(t) \delta(t - \tau_p(t))
\label{eq:3}
\end{align}
where $P$ is the number of resolvable paths, $h_p(t)$ is the $p$-th path's amplitude, and $\tau_p(t)$ is the $p$-th path's delay. 

Each path's amplitude and delay are assumed to be time-invariant for several symbol transmissions. Hence, the received baseband signal after carrier demodulation is given by
\begin{align}
y(t) =  \sum_{n=0}^{N-1}\sum_{p=0}^{P-1} {\widetilde h}_p \psi^{(n)}(t- nT-\tau_p) + n(t)
\label{eq:4}
\end{align}
where ${\widetilde h}_p \triangleq \sqrt{E}h_p e^{-j2\pi f_c \tau_p} \in \mathbb{C}$ is the $p$-th path's energy-including channel coefficient, $\tau_p$ is the $p$-th path's delay, and $n(t)$ denotes additive noise. 

A PN-training block consisting of antipodal bits $\{\pm 1\}^{N_{pn}}$ precede chirp symbol block for synchronization and channel estimation purposes. A guard interval $T_g$ is inserted between the PN-training and chirp symbol block for avoiding inter block interferences. In addition, we employ gray code for bit mapping of adjacent chirp waveforms for reducing detection errors. 

Our problem objective is to design $M$-ary orthogonal chirp waveforms for coherent and non-coherent receivers for UW-A communications. In the coherent receiver, channel estimation and equalization is required, whereas an optimal non-coherent detection simply match-filters received signals with orthogonal chirp waveforms.

\vspace{-10pt}
\section{Receiver Design}
\label{s:3}

Receiver design considerations for a coherent and a non-coherent receivers are presented. For both coherent and non-coherent detectors, PN-training sequences are utilized for packet synchronization. Then, coherent receiver implements channel estimation, chirp matched filtering, and symbol detection techniques. On the other hand, a non-coherent receiver only conducts matched filtering.

\subsection{Coherent Receiver}
We first sample the received baseband signal over the total frame duration $NT$ at the sampling frequency $f_s \triangleq \frac{1}{T_s}$. Each symbol consists of $L \triangleq \frac{T}{T_s}$ number of samples. Then, multipath pulses are assumed to extend duration of $L_P = L + P-1$ pulses. The $n$-th received baseband vector is written as
\begin{align}
\mathbf{y}_n = \mathbf{H} \bm{\psi}^{(n)} + {\mathbf{n}}_n ~\in {\mathbb{C}}^{L_P}, \quad n= 0, \dots, N-1
\label{eq:5}
\end{align}
where $\mathbf{H}$ is the multipath channel matrix

\begin{align}
{\mathbf H} \triangleq \left[\begin{array}{ccccc}
       h_0 & 0           & \cdots & 0       & 0 \\
       h_1 & h_0         & \cdots & 0       & 0   \\
       \vdots &\vdots    & \vdots & \vdots  & \vdots\\
       h_{P-1} & h_{P-2} &        & 0       & 0  \\
       0    & h_{P-1}    &        & h_0       & 0 \\
       \vdots &\vdots    & \vdots & \vdots  & \vdots\\
       0    & 0          & \cdots & h_{P-1} & h_{P-2} \\
       0    & 0          & \cdots & 0       & h_{P-1}
     \end{array}\right] \in \mathbb{C}^{L_P\times L}
\label{eq:6}
\end{align}
$\mathbf{y}_n = \begin{bmatrix} y((n-1)T), \dots, y(nT+(P-2)T_s) \end{bmatrix} ^T$ is the $n$-th received baseband vector, $\bm{\psi}^{(n)} \in \mathbb{C}^{L}$ is the $n$-th  chirp vector, and $\mathbf{n}_n \in \mathbb{C}^{L_P}$ is the additive noise.

Channel estimation is conducted by a priori known PN-training sequences. Then, we consider $N_P = N_{pn} + P - 1$ extended pulses, so the received PN-training vector $\mathbf{y}_{tr}$ is given by
\begin{align}
\mathbf{y}_{tr} = \mathbf{B}_{tr} \mathbf{h} + \mathbf{n}_{tr} ~\in {\mathbb{C}}^{N_P}
\label{eq:7}
\end{align}
where $\mathbf{B}_{tr}$ is a PN-training Toeplitz matrix

\begin{align}
{\mathbf B}_{tr} \triangleq \left[\begin{array}{ccccc}
       b_0    & &            &    &  \bf 0 \\
              & & \ddots     &    &    \\
       \vdots & &            &    & b_0\\
       b_{N_{pn}-1}    & &            &    & \vdots    \\
              & & \ddots     &    &         \\
       \bf 0  & &            &    & b_{N_{pn}-1}
     \end{array}\right] \in \mathbb{C}^{N_P \times P}
\label{eq:8}
\end{align}
$b_i \in \{\pm 1\}_{i=0}^{N_{pn}-1}$, $\mathbf{h} = [h_0,\cdots, h_{P-1}]^T \in {\mathbb{C}}^{P}$ is the multipath channel vector, and $\mathbf{n}_{tr} \in \mathbb{C}^{N_P}$ is the noise vector.

Then, multipath channel vector can be solved by the following least-squares (LS) problem
\vspace{2pt}
\begin{align}
\mathbf{\widehat{h}} = \it {arg} \min_{\mathbf{h} \in \mathbb{C}^{P}} \left | \left |\mathbf{y}_{tr} - \mathbf{B}_{tr} \mathbf{h}\right | \right |_{\mathrm 2}^{\mathrm 2}
\label{eq:9}
\end{align}
where $\|\cdot\|_2$ denotes the Euclidean norm. For a given PN-training matrix $\mathbf{B}_{tr}$, the solution to (\ref{eq:9}) is
\vspace{2pt}
\begin{align}
\mathbf{\widehat{h}} = (\mathbf{B}_{tr}^H\mathbf{B}_{tr})^{-1}\mathbf{B}_{tr}^H\mathbf{y}_{tr}
\label{eq:10}
\end{align}
Under additive white Gaussian noise (AWGN) condition, (\ref{eq:10}) is the maximum-likelihood (ML) optimal channel estimate \cite{Proakis07}.

Based on the estimated channel matrix $\mathbf{\widehat{H}}$, the decision criterion for the $n$-th transmitted symbol is the chirp vector that maximizes the correlation of the chirp vector and the $n$-th channel-equalized received vector represented as
\begin{equation}
\label{eq:11}
\begin{aligned}
\widehat{\bm{\psi}}^{(n)} \triangleq \arg \max_{\bm{\psi}^{(n)} \in \left\{\bm{\psi}_0,\cdots,\bm{\psi}_{M-1}\right\}} \left | \left | (\bm{\psi}^{(n)})^H (\widehat{\mathbf{H}}^H \widehat{\mathbf{H}})^{-1} \widehat{\mathbf{H}}^H\mathbf{y}_n \right | \right |
\end{aligned}
\end{equation}

\subsection{Non-coherent Receiver}
\begin{figure}
  \centering
  \includegraphics[width=0.7\textwidth]{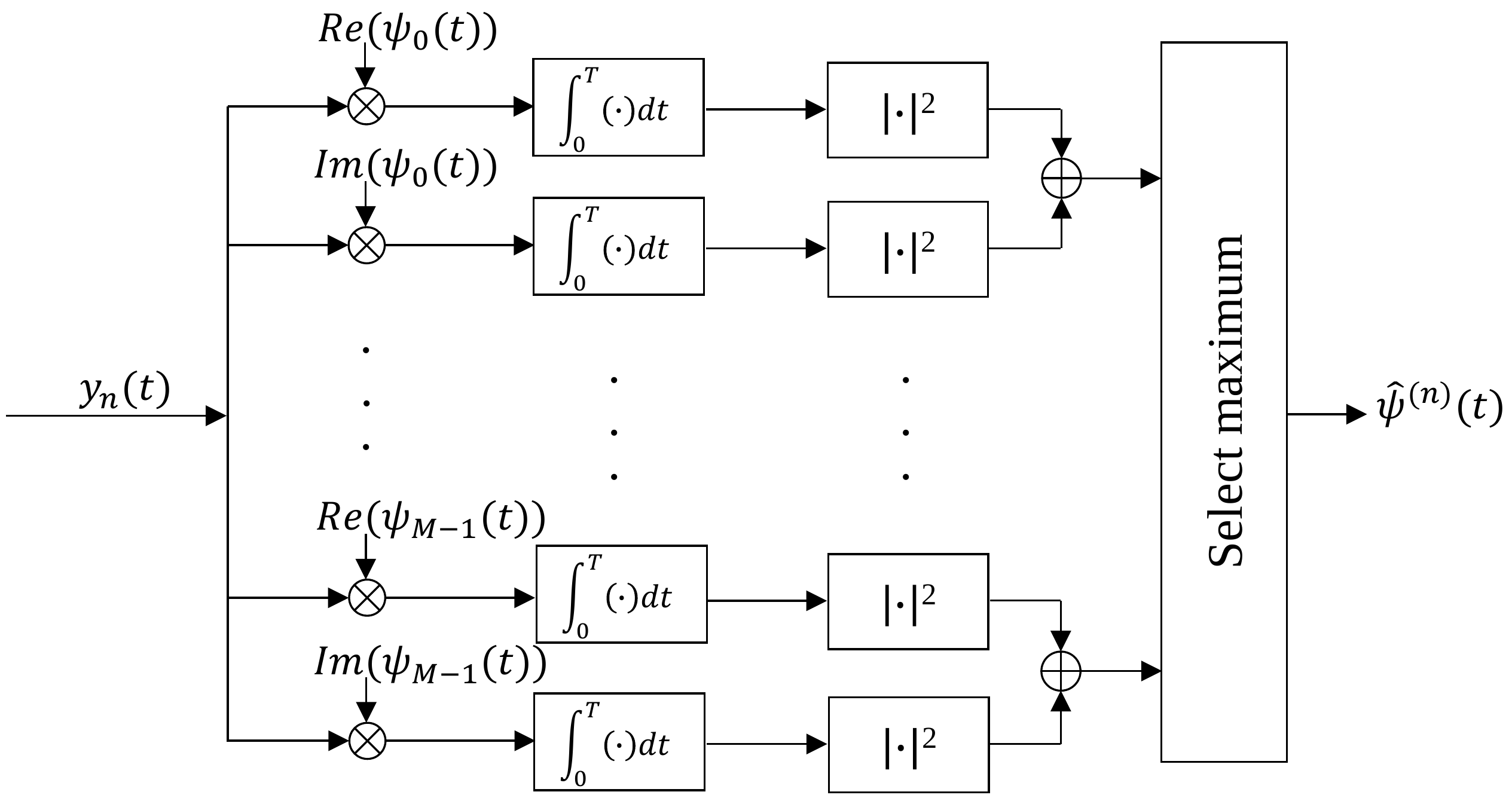}
  \caption{Optimal non-coherent receiver.}
  \label{fig1}
\end{figure}

Since proposed chirp waveforms are orthogonal, an optimal non-coherent receiver \cite{Proakis07} is designed as Fig. \ref{fig1}. Then, the optimal non-coherent receiver in vector form can be represented as
\vspace{2pt}
\begin{equation}
\label{non_det_vec} 
\begin{aligned}
\widehat{\bm{\psi}}^{(n)}= \it {arg} \max_{\bm{\psi}^{(n)} \in \left\{\bm{\psi}_0,\cdots,\bm{\psi}_{M-1}\right\}} \left |Re(\bm{\psi}^{(n)})^T \widetilde{\mathbf{y}}_n \right |^{\mathrm 2} + \left |Im(\bm{\psi}^{(n)})^T \widetilde{\mathbf{y}}_n \right |^{\mathrm 2}
\end{aligned}
\vspace{2pt}
\end{equation}
\normalsize
where $\mathbf{\widetilde{y}}_n = \begin{bmatrix} y((n-1)T), \dots, y(nT-T_s) \end{bmatrix} ^T \in {\mathbb{C}}^{L}$ is the $n$-th received vector without extended pulses. The received vector $\mathbf{\widetilde{y}}_n$ match-filters with sampled chirp vectors, respectively. Then, the transmitted chirp vector is decided as the one that maximizes the correlation output.

In the optimal non-coherent receiver, none of any channel estimation or equalization technique is exploited, which implementations become easier. However, the performance of the non-coherent receiver is expected to deteriorate due to lacking of CSI in symbol detections. 

\vspace{-2.5pt}

\section{Orthogonal Chirp Waveform Design}
\label{s:4}
The characteristics of the orthogonality of proposed chirp waveforms are analyzed by their cross-correlation coefficients. In addition, theoretical BER values for coherent and non-coherent receivers are derived explicitly. 

\subsection{Cross-correlation Coefficient}
The cross-correlation coefficient $\rho_{kl}$ of the $k$-th and $l$-th $(0\leq k,l\leq M-1,~k\neq l)$ chirp waveforms is derived as
\begin{equation}
\begin{aligned}
\rho_{kl}  &= \int_0^{T} \psi_{k}(t) \psi_{l}^*(t)dt\\
      &= \frac{1}{T}\int_0^{T} e^{j2\pi(k-l)\Delta ft}dt\\
      &= \frac{1}{T}\left[\frac{e^{j2\pi(k-l)\Delta f T}-1}{j2\pi(k-l)\Delta f}\right]\\
      &= 0
\end{aligned}
\normalsize
\label{eq:13}
\end{equation}
where $\Delta f$ are chosen to be multiples of $\frac{1}{T}$ for guaranteeing orthogonality. The rationale behind the parameters' selection criteria is: for a coherent receiver, $\Delta f$ is select to be multiples of $\frac{1}{2T}$, whereas for non-coherent detection, $\Delta f$ is chosen to be multiples of $\frac{1}{T}$ \cite{Proakis07}. Thus, for satisfying both constraints for coherent and non-coherent detections, $\Delta f$ should be chosen as multiples of $\frac{1}{T}$. Since all the cross-correlation coefficients between different chirp waveforms are zeros, so proposed chirp waveforms constitute a $M$-ary orthogonal chirp modulation. 

\subsection{BER Analysis}
On account of the orthogonality of proposed chirp waveforms, BER of orthogonal chirp modulation for coherent detection in AWGN \cite{springer00} is given by
\begin{align}
P_{be,co-OCK} = Q\left(\sqrt{\frac{E}{\log_2 M \cdot N_0}}\right)
\label{eq:14}
\end{align}
where $E$ is the energy per symbol, $\log_2 M$ is the number of bits per symbol, and $\frac{N_0}{2}$ is the variance of additive noise. 

On the other hand, BER of the optimal non-coherent detector in AWGN \cite{Proakis07} is written as
\begin{align}
P_{be,non-OCK} = \frac{1}{2}\exp\left(-\frac{E}{2\log_2 M \cdot N_0}\right)
\label{eq:15}
\end{align}

Theoretical BER values for coherent and non-coherent receivers in AWGN are analyzed and represented in explicit equations. Then, we can evaluate the performance of orthogonal chirp waveforms.

\section{Experimental Studies}
\label{s:5}
\begin{figure}
 \centering
  \includegraphics[width=0.7\textwidth]{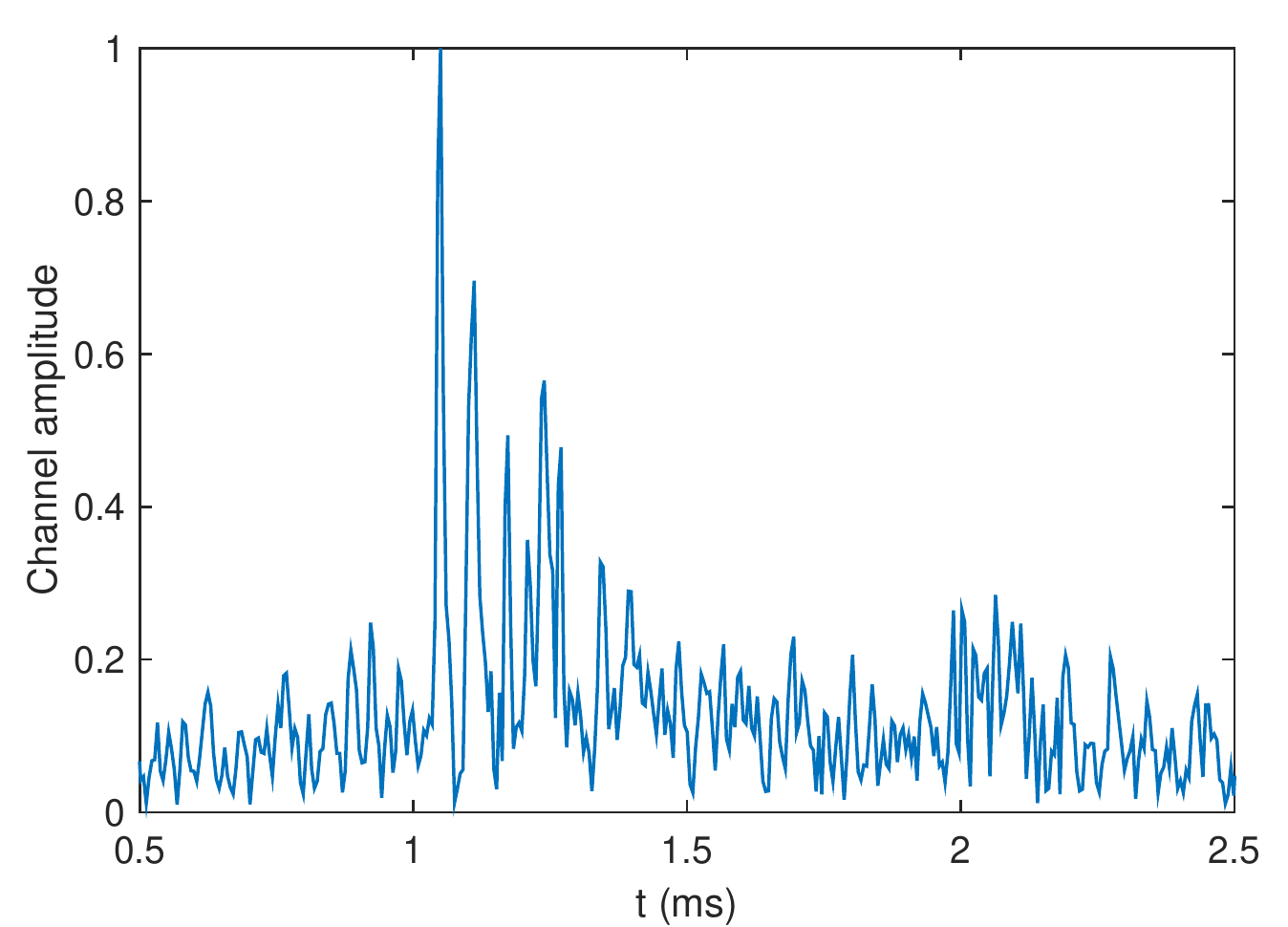}
  \caption{Normalized channel response in tank experiments.}
  \label{fig_ch}
\end{figure}

Performance of proposed orthogonal chirp waveforms is evaluated in indoor water tank experiments implemented by software-defined radio platforms. Two USRP-N210s are interfaced with Teledyne RESON TC4013 omnidirectional hydrophones, operating in frequency range $0 - 170$ kHz. The hydrophones are deployed in an indoor water tank with dimensions $8$ ft $\times$ $2.5$ ft $\times$ $2$ ft. Distance between two hydrophones is $d$ = $1.02$ m. Parameters are set as: the symbol duration $T$ = $0.33$ ms, the PN-training duration $T_{tr}$ = $1.31$ ms, the guard period $T_g$ = $5.12$ ms, the initial frequency $f_0$ = $3.05$ kHz, the frequency spacing $\Delta f$ = $3.05$ kHz, the chirp rate $\mu$ = $9.31\times 10^6$ Hz/s, and carrier frequency $f_c$ = $100$ kHz. Each transmitted packet contains $N$ = $32$ information symbols and over $3000$ packets are recorded. Normalized channel response for one realization is shown in Fig. \ref{fig_ch}, number of resolvable paths $P$ = $4$.

\begin{figure}
 \centering
  \includegraphics[width=0.7\textwidth]{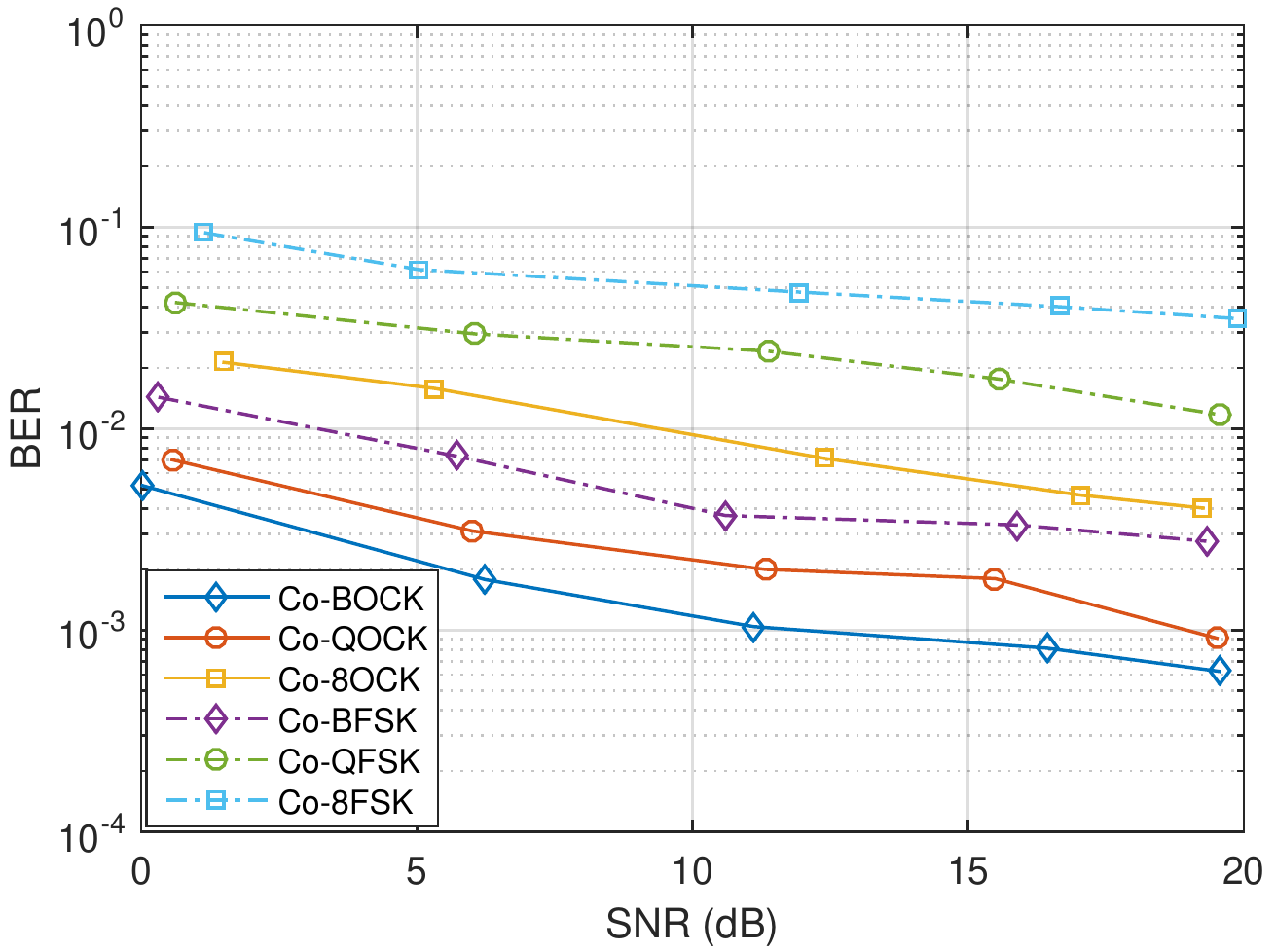}
  \caption{BER for the coherent receiver in tank experiments.}
  \label{fig_exp_co_exp}
\end{figure} 

\begin{figure}
 \centering
  \includegraphics[width=0.7\textwidth]{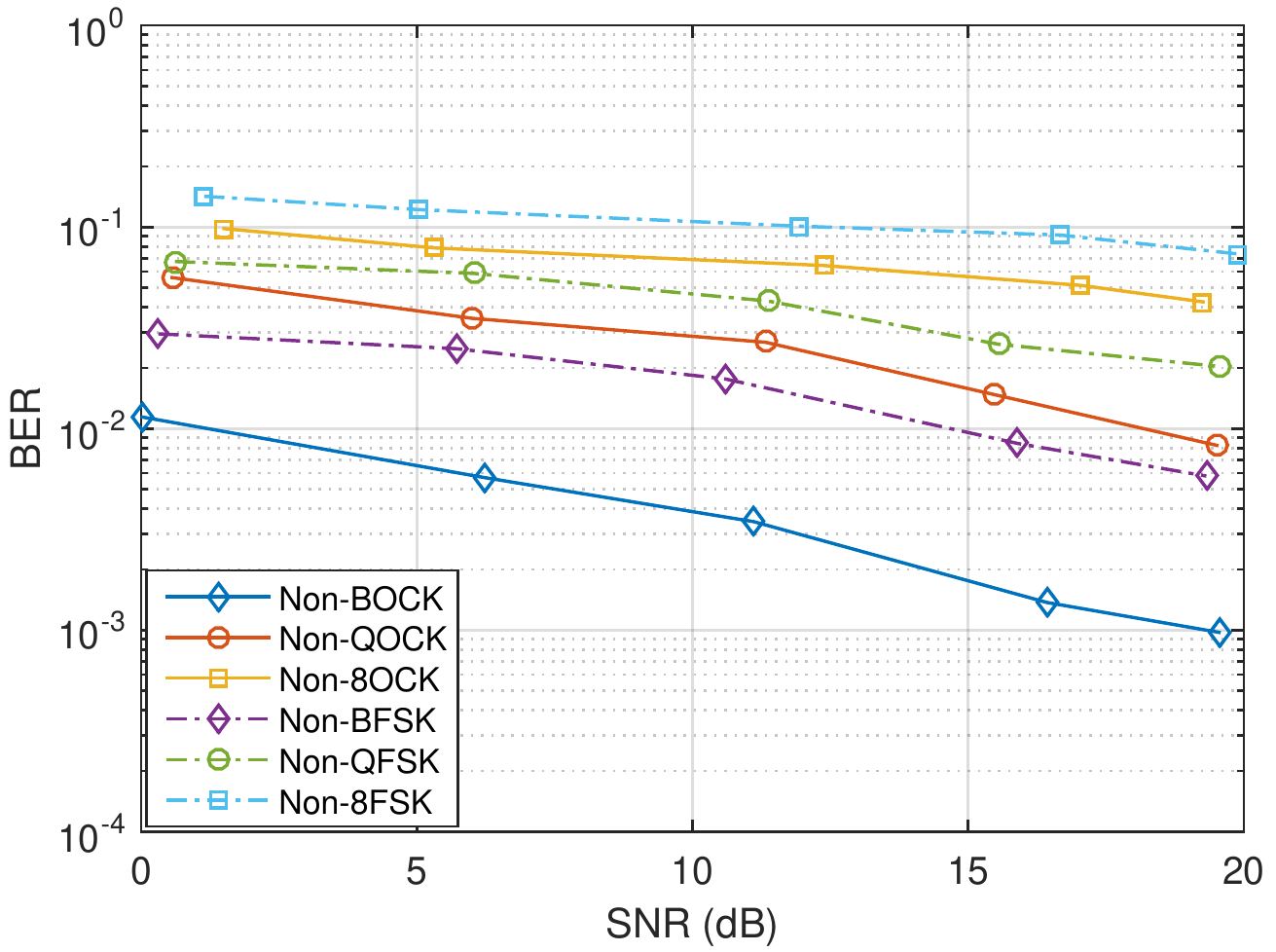}
  \caption{BER for the non-coherent receiver in tank experiments.}
  \label{fig_exp_non_exp}
\end{figure} 


Fig. \ref{fig_exp_co_exp} shows the performance of the coherent detections in tank experiments. We discover that the higher the order of the modulation, the higher the BER values due to decreasing energy per bit (energy per symbol is fixed for fairness of comparison). If comparing among binary modulations, proposed BOCK improves performance for $6.11$ dB than that of BFSK at SNR = $15.49$ dB. At SNR = $6.01$ dB, BER of 8-OCK increases for $8.85$ dB than that of QOCK due to closer Euclidean distance between constellation symbols. At SNR = $19.26$ dB, BER of coherent 8-OCK is $3.43 \times 10^{-3}$ and bit rate can achieve up to $9.16$ kbit/s. 


Fig. \ref{fig_exp_non_exp} depicts the performance of optimal non-coherent receivers in tank experiments. It demonstrates that at SNR = $10.62$ dB, BER of BOCK has a $7.83$ dB decrease than that of BFSK. Then, at SNR = $15.49$ dB, 8-OCK increases BER for $5.51$ dB than that of QOCK, whereas the bit rate is $1.5$ times. Hence, we realize that there is a trade-off between BER and the bit rate when designing communication systems. The optimal non-coherent 8-OCK can achieve BER = $4.25 \times 10^{-2}$ at SNR = $19.26$ dB.
 
\section{Conclusion}
\label{s:6}

In this paper, we propose a $M$-ary orthogonal chirp modulation in underwater acoustic multipath channel. Cross-correlation coefficients and theoretical BER values of coherent and optimal non-coherent detectors are derived in closed-form expressions. Moreover, BER performance of proposed chirp waveforms is evaluated by in-house built software-defined acoustic modems in indoor tank testbeds. Experimental results demonstrate that orthogonal chirp waveforms can improve BER performance than that of FSK significantly in both coherent and non-coherent receiver designs. Therefore, proposed $M$-ary orthogonal chirp modulation can enhance effectiveness of communication systems in underwater acoustic environments.


\bibliographystyle{IEEEtran}
\bibliography{ref}

\end{document}